\documentclass[conference]{IEEEtran}

\usepackage{enumerate}
\usepackage{epsfig}
\usepackage{graphicx}
\usepackage{amsmath}
\usepackage{amssymb}
\usepackage{algorithm}
\usepackage[noend]{algpseudocode}
\usepackage{bm}
\usepackage{url}
\usepackage{scrextend}
\usepackage{listings}  
\usepackage{subfigure}
\usepackage{amsmath}
\usepackage{rotating}
\usepackage{wrapfig}
\usepackage{multirow}
\usepackage{lscape}
\usepackage{verbatim}
\usepackage{wrapfig}
\usepackage{hhline}
\usepackage{textcomp,booktabs}
\usepackage{xcolor}
\usepackage{paralist}
\usepackage{cite}
\usepackage{color, colortbl}
\usepackage{float}

\definecolor{Gray}{gray}{0.7}

\def\BState{\State\hskip-\ALG@thistlm}

\begin{document}
	
	\title{Attacks and Defenses in Mobile IP: Modeling with Stochastic Game Petri Net}
	
	\author{
		\IEEEauthorblockN{Sajedul Talukder}
		\IEEEauthorblockA{Florida Int'l University\\
			Florida, USA\\
			stalu001@cs.fiu.edu}
		\and
		\IEEEauthorblockN{Md. Iftekharul Islam Sakib}
		\IEEEauthorblockA{BUET\\
			Dhaka, Bangladesh\\
			miisakib@gmail.com}
		\and
		\IEEEauthorblockN{Md. Faruk Hossen}
		\IEEEauthorblockA{BUET\\
			Dhaka, Bangladesh\\
			faruk.08.cse@gmail.com}
		\and
		\IEEEauthorblockN{Md. Shohrab Hossain}
		\IEEEauthorblockA{BUET\\
			Dhaka, Bangladesh\\
			mshohrabhossain@cse.buet.ac.bd}
	}
	
	\maketitle
	
	\begin{abstract}
		The urging need for seamless connectivity in mobile environment has contributed to the rapid expansion of Mobile IP. Mobile IP uses wireless transmission medium, thereby making it subject to many security threats during various phases of route optimization. Modeling Mobile IP attacks reasonably and efficiently is the basis for defending against those attacks, which requires quantitative analysis and modeling approaches for expressing threat propagation in Mobile IP. In this Paper, we present four well-known Mobile IP attacks, such as Denial-of-Service (DoS) attack, bombing attack, redirection attack and replay attack and model them with Stochastic Game Petri Net (SGPN). Furthermore, we propose mixed strategy based defense strategies for the aforementioned attacks and model them with SGPN. Finally, we calculate the Nash Equilibrium of the attacker-defender game and thereby obtain the steady state probability of the vulnerable attack states. We show that, under the optimal strategy, an IDS needs to remain active 72.4\%, 70\%, 68.4\% and 66.6\% of the time to restrict the attacker's success rate to 8.5\%, 6.4\%, 7.2\% and 8.3\% respectively for the aforementioned attacks, thus performing better than the state-of-the-art approach.
	\end{abstract}
	
	\section{\textbf{Introduction}} \label{sec:intro}
	
	Wireless communication has witnessed a massive growth in number of users in the recent years. Spreading of wireless networks has influenced everyday life, from e-governance~\cite{TSRICIEV14} to social networks~\cite{TC2017}, from digital automation~\cite{TSRICEEICT14} spreading up to space and aeronautical networks~\cite{931283}. One of the key benefits of wireless technology is mobility, which allows mobile users to move from one network to another while maintaining their home IP address unchanged~\cite{so2006mobile}. Mobile IP (RFC 2002) is a standard protocol established by the Internet Engineering Task Force (IETF) that builds on the Internet Protocol by making mobility transparent to applications and higher level protocols like TCP. Mobile IP settings mostly exist in wireless networks where users need to carry their devices across several networks with different IP address. 3G and 4G networks also use Mobile IP to provide transparency when user of the internet migrates between cellular towers~\cite{song2007interworking}. Mobile IP allows the mobile node to use two IP addresses: a fixed home address and a care-of address that changes at each new point of attachment. Now a days, a vast number of devices are facilitated with Mobile IP that includes Cellular phones, PDAs, GPS, Tablets and other handheld devices. 
	More and more entities getting connected to the wireless networks, the security threats that causes massive impairment are bulging as well. Abundant of malicious nodes that have been thrust into networks across the world have made the detection of Mobile IP attacks more difficult. Network security, in particular Mobile IP security needs to provide confidentiality, availability and integrity for data communication. 
	However, the need to provide unbroken session as the user or node moves from one link to another without human intervention and non-interactivity has created the scope for the attackers to perform various attacks in Mobile IP. Our study focuses on understanding and modeling the following Mobile IP attacks and their appropriate defense strategies: (1) Denial of Service attack, (2) Redirection attack, (3) Replay attack and (4) Bombing attack. All of the attacks are mostly due to route optimization between the Mobile Node and Corresponding Node. Mobile Node that changes its IP address needs to update its care-of-address and send the binding update to the Corresponding Node. Binding updates are vulnerable to various attacks since Malicious Node can penetrate the route between Mobile Node and Corresponding Node. Attacker can steal information, alter it or redirect it by fooling either Corresponding Node or Mobile Node or both. 
	Researchers have tried to analyze Mobile IP attacks and their defenses using various approaches like using IPSec~\cite{yokote2001method}, number of independent data networks~\cite{haverinen2002ip}, nondisclosure method~\cite{fasbender1996variable}, IP security primitives~\cite{inoue1997secure, braun2001secure}, authenticating a mobile node~\cite{leung2004mobile}, using public-key~\cite{zao1999public}, security association policy server~\cite{yokote2002intelligent} or securing binding update~\cite{deng2002defending}. However, none of the above mentioned works attempted to model and analyze the attacks and defense scenarios using Stochastic Game Petri Net (SGPN). To the best of our knowledge, this paper is the first attempt to model attacks and defenses in Mobile IP using SGPN. 
	
	\noindent
	{\bf Our Contributions}.
	This paper presents the following contributions:
	
	\begin{compactitem}
		
		\item
		{\bf Attack Analysis}.
		Analyze Mobile IP and four major attacks such as Denial-of-Service (DoS) attack, bombing attack, redirection attack and replay attack present in Mobile IP.
		
		\item
		{\bf Attack Modeling}.
		Model the attacks using stochastic game petri net (SGPN).
		
		\item
		{\bf Defense Modeling}.
		Propose mixed strategy based defense strategies for the aforementioned attacks and model them with SGPN.
		
		\item
		{\bf Evaluation}.
		Evaluate the models by calculating the steady state probability of the vulnerable attack states and show that, an IDS needs to remain active 72.4\%, 70\%, 68.4\% and 66.6\% of the time to restrict the attacker's success rate to 8.5\%, 6.4\%, 7.2\% and 8.3\% respectively for the aforementioned attacks. .  
		
	\end{compactitem}
	
	The rest of the paper is organized as follows. Section II describes the background of the work. Section III analyzes the previous works. 
	Section IV illustrates the attacks modeling and our proposed solutions to the attacks, Section V deals with the findings and result analysis and finally Section VI concludes the paper with a highlight on the scope of future work.
	
	\section{\textbf{Background}} \label{sec:background}
	
	\subsection{\textbf{Mobile IP}}
	IP address is obtained from a TCP connection which uniquely identities a device/node's point of attachment to the internet. When a device moves from its home network to a new foreign network, it needs to change its IP address and re-establish a new TCP connection. If there is an ongoing session going on with this device, the session needs to be disconnected until a new IP address of a moving device is obtained. Mobile IP Protocol was proposed by a working group within the Internet Engineering Task Force (IETF) in order to solve this mobility issue. Mobile IP for IPv4 is described in IETF RFC 5944, and extensions are defined in IETF RFC 4721. Mobile IPv6, the IP mobility implementation for the next generation of the Internet Protocol, IPv6, is described in RFC 6275.
	
	\begin{figure}
		\begin{center}
			\includegraphics[width=\columnwidth]{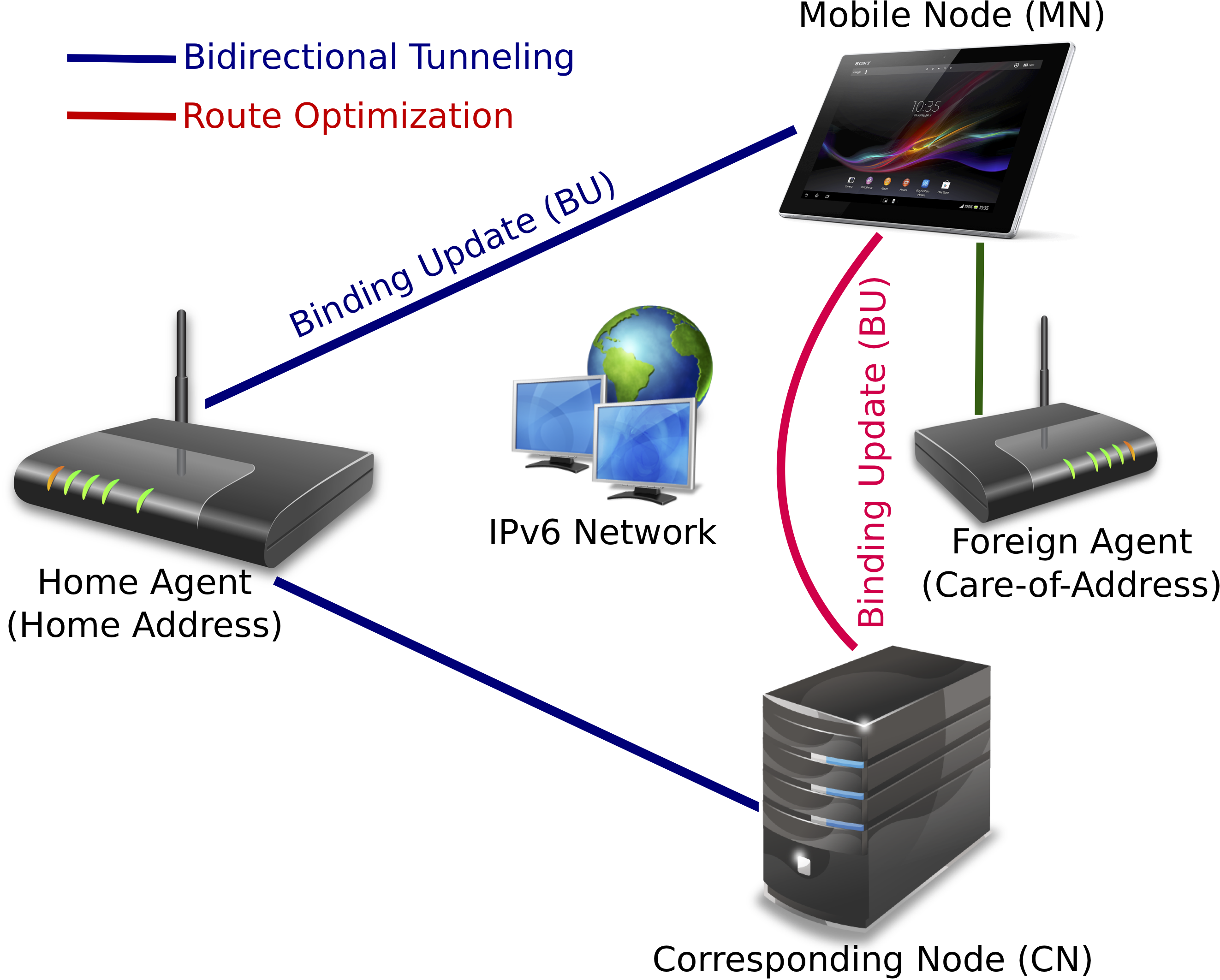}
			\caption{MIPv6 Architecture}
			\label{fig:architecture}
		\end{center}
	\end{figure}
	
	\subsubsection{\textbf{Mobile IP Overview}}
	Mobile IP designed by IETF is a standard protocol to enable mobile users to move across the network while maintaining their permanent IP address. IP datagram can be routed over Internet transparently using Mobile IP. Each mobile node has a Home Address (HoA) that corresponds to its home network that remains unchanged regardless of its current location. If a node moves to foreign network, it is associated with a Care-of Address (CoA), that provides information of its current location. Using Mobile IP, the device can change its location across various networks without requiring to change its IP address. Home Agent (HA) and Foreign Agent (FA) are the routers that perform the encapsulation and decapsulation of the datagram packets respectively. 
	\subsubsection{\textbf{Mobile IP Terminologies}}
	Now we present the most common terminologies used in Mobile IP.
	\begin{itemize}
		\item \textbf{Mobile Node (MN):} A moving device that connects to the internet using a fixed home address and can change the location and point of attachment to the internet while keeping ongoing communication uninterrupted. Examples of MNs are Cellular phones, PDAs, GPS, tablets, routers and other handheld devices.
		\item \textbf{Home Network (HN):} Network within which a MN identifies its home address.
		\item \textbf{Home Address (HoA):} An IP address assigned to MN for a home network that remains same regardless of where the device is attached to the internet. 
		\item \textbf{Home Agent (HA):} A router on the MN's home network that tracks the MN's current location (CoA), intercept, encapsulates and tunnels datagram packets to the MN when it is away from home.
		\item \textbf{Foreign Network (FN):} Any network other than the MN's home network, on which MN moves its point of attachment.
		\item \textbf{Care-of-Address (CoA):} A temporary IP address assigned to a MN while it is visiting a foreign network away from its home network. 
		\item \textbf{Foreign Agent (FA):} A router on the MN's foreign network that provides a CoA to the MN and acts as a default router for datagram generated by the MN. It also decapsulates and delivers datagram to the MN that are encapsulated by the MN's HA.
		\item \textbf{Correspondent Node (CN):} A mobile or stationary device that sends or receives packets to or from the MN. 
		\item \textbf{Binding Update (BU):} Message used to notify the HA or CN about the current location of the MN by sending the CoA.
	\end{itemize}
	
	\subsubsection{\textbf{Mobile IPv6}}
	In MIPv6, the Mobile Node (MN) can communicate in two ways with the Corresponding Node (CN), through bidirectional tunneling and through route optimization, see figure~\ref{fig:architecture}. In bidirectional tunneling, packets from the CN are sent to the HA, which forwards them to the MN through a tunnel. The MN sends the responses through a reverse tunnel to the HA, which forwards the data to the CN. Communication between MN and CN thus always happens via HA. Whenever, MN changes its network and moves to a new network, it gets a new CoA. MN needs to notify HA about its new CoA and this is done through Binding Update (BU).
	
	\textbf{Route Optimization.} It is possible to bypass the requirement of HA for every packet transfer through the use of route optimization. When route optimization is enabled, any communication after the initial connection through the HA is handled directly between the MN and the CN. A type 2 routing header is used for this process. Since communication occurs directly between MN and CN instead of detouring via the HA, it becomes more efficient because routing can be optimized. However, in order to use route optimization, CN must have Mobile IPv6 support on it. Similar to bidirectional tunneling, MN needs to notify CN about its new CoA through BU. Route optimization is one of the main benefits of Mobile IPv6 over Mobile IPv4.
	
	\textbf{Return Routability.} In order to prevent redirection attacks during the exchange of BU between MN and CN, an authentication test is needed. This authentication process is called the return routability procedure and lets the CN test whether the MN is actually reachable via both its CoA and its HoA. If the authentication is successful, only then are BU accepted by the CN.

	\subsection{\textbf{Stochastic Game}}
	A stochastic game is a n-player game in which players’ payoff and the probability distribution of a new state being visited depend on the collection of actions that the players choose, together with the current state. A Stochastic Game is represented as a 7-tuple vector $SG = (I, X, A_i, A, r_i, q, \beta),$ where
	\begin{addmargin}[1.5em]{0em}
		(1) $I = \left\{1, 2, . . . , n\right\}$ is the set of players;\\
		(2) $X = \left\{1, 2, . . . , N\right\}$ is the state space;\\
		(3) $A_i$ is the finite action set of player $i$;\\
		(4) $A := A_1 \times ...\times A_n$ is the set of action profiles;\\
		(5) $r_i : X\times A \rightarrow R$ is the reward function of a player $i$;\\
		(6) $q(x\prime\mid x,a)$ is the probability of next state being $x\prime$ given that the current state is x and the action profile $a \in A$;\\
		(7) $\beta \in (0,1)$ is a discount factor.\\
	\end{addmargin}
	
	\subsection{\textbf{Petri Net}}
	
	Petri net (a.k.a. place/transition net) is a mathematical and graphical modeling language for the description and analysis of concurrent processes in distributed systems. A Petri net is a directed bipartite graph consisting of places, transitions and arcs. Places denote conditions which are represented by circles. Transitions denote events which are represented by rectangular bars. Arcs run from a place to a transition or vice versa. Arc never runs between places or between transitions. Petri Net is very similar to State Transition Diagrams. Graphically, places in a Petri net may contain a discrete number of tokens. Marking M is the state of the net at any time in terms of the distribution of tokens over the places. Transitions may be fired to transfer control from input place to output place through the exchange of tokens. A firing is an atomic event.\\

	\section{\textbf{Stochastic Game Petri Net (SGPN)}}
	
	We now present the formal representation of Stochastic Game Petri Net (SGPN) which is derived from\cite{wang2013modeling}:
	
	\textbf{SGPN.} A SGPN is represented as a 9-tuple vector $SGPN = (N, P, T, F, \pi, \lambda, R, U, M_{0}),$ where
	\begin{addmargin}[1.5em]{0em}
		(1) N = {1, 2, . . . , n} is the set of players;\\
		(2) P is a finite set of places;\\
		(3) T $= T^{1} \cup T^{2} \cup . . . \cup T^{N}$ is a finite set of transitions, where $T^{k}$ is the set of transitions with respect to player k for $k \in N;$\\
		(4)  $\pi : T \rightarrow [0,1]$ is a routing policy representing the probability of choosing a particular transition;\\
		(5) $F \subseteq I \cup O$ is a set of arcs, where $I \subseteq P  \times T$ and $O \subseteq T \times P$ such that $P \cap T = \phi$ and $P \cup T  \neq \phi ,$ where $\phi$ is an empty set; we denote $x = \left\{y \mid (y,x) \in F\right\} $ the preset of x, similarly, $x = \left\{y \mid (x,y) \in\\ F\right\}$ the post-set of x;\\
		(6) $R : T \rightarrow (R_{1}, R_{2}, . . . , R_{n})$ is a reward function for the players taking each transition, where $R_{i} \in (-\infty, +\infty), i \in N;$\\
		(7) $\lambda = \left\{\lambda_{1}, \lambda_{2}, . . ., \lambda_{w}\right\}$ is a set of transition firing rates in the transition set, where w is the number of transitions;\\
		(8) U is the utility function of players; and\\
		(9) $M_{0}$ is the initial marking.
	\end{addmargin}
	
	Each token $S$ is assigned a reward vector $h(s) = (h_1(s), h_2(s),. . ., h_n(s))$ as its property, where $h_k(s)$ is the reward of player
	$k$ in token $s$. Players get the reward $R(t)$ after the firing of the transition $t$, and the reward is recorded in the reward vector $h$ of the token\cite{wang2013modeling}. For the sake of simplicity and to fit our attacker-defender model, we assume that our stochastic game is a two-player discounted stochastic game.\\
	
	\textbf{Definition 1 (Transition Probability Matrix).} Given a SGPN with $r$ places, a Transition Probability Matrix is a $r \times r$ matrix M, where $M_{ij}$ represents the probability of a transition $t_k$ being fired such that $p_i$ is the input place, $p_j$ is the output place and $p_i,p_j \in P, t_k \in T$.\\
	
	\textbf{Definition 2 (Strategy).} Given a player $k$ in SGPN, strategy is a vector $\pi^k = (\pi(t_1^k), \pi(t_2^k),...,\pi(t_{wj}^k))$, where $\pi(t_j^k)$ is the probability that player $k$ takes action $t_j$ and $w_j = \mid{T^k}\mid$.\\
	
	\textbf{Definition 3 (Utility).} Given a SGPN in infinite time horizon with $n$ players, utility of a player $i$ is $U^i(p,\pi) = \sum^\infty_{k=1} \beta^{k-1} R_i^k(p,\pi)$, where $\pi = (\pi_1, \pi_2,...,\pi_n)$, $\beta \in (0,1)$ is a discount factor and $R_i^k(p,\pi)$ is the expected reward function of player $i$ at the $k$-th stage of the game. It can be further simplified to $U^i(\pi)$ when $p$ denotes the initial state of player $i$.\\
	
	\textbf{Definition 4 (Nash Equilibrium).} Given a SGPN, a Nash Equilibrium  for a two-player stochastic game is a vector profile $\pi^* = (\pi_1^{*}, \pi_2^{*},...,\pi_n^{*})$ such that $U^i(p,\pi^{*}) \geq U^i(p, \pi_i, \pi_{-i}^{*})$ for all $p \in P, i \in N$, where $\pi_{-i}^{*}$ is any alternative mixed strategy
	of player $i$ except $\pi^{*}$.\\
	
	\textbf{Thoerem 1.} Every discounted stochastic game is guaranteed to have a Stationary Nash Equilibrium.\\
	\textbf{\textit {Proof.}} Let, $F^i = (f^i_1,f^i_2,...f^i_n)$ be the set of stationary strategies for player $i$.
	
	A Stationary Nash Equilibrium belongs to the class of strategy profiles $F^1 \times...\times F^n$. We need to prove that, there exists $f^* = (f^*_1,f^*_2,...f^*_n) \in F:= F^1 \times...\times F^n$ such that $U^i(p,f^*) \geq U^i(p,\pi_i,f^*_{-i})$ for all $p \in P, i \in N$.
	
	We note that $F$ is a compact convex set in some Euclidean space. We also observe that $U^i(p,.)$ is continuous on $F$.
	
	We get,
	\begin{addmargin}[2.5em]{0em}
		$(f^k_1,f^k_2,...f^k_n) \rightarrow (f_1,f_2,...f_n)$ as $k \rightarrow \infty$
	\end{addmargin}
	$\implies U^i(p,f_1^k,...,f_n^k) \rightarrow U^i(p,f_1,...,f_n)$     (1)
	\begin{addmargin}[2.5em]{0em}
		Equation (1) follows directly from the formula:\\
		$U^i(f_1,...,f_n) = {(N -\beta Q_{f_1...f_n})}^{-1}R_i(f_1,...,f_n)$ where $Q$ denotes the probability of the next state given the current state.
	\end{addmargin}
	
	\textbf{\textit {Claim.}} A policy $f^* = (f^*_1,f^*_2,...f^*_n)$ is a Nash Equilibrium if $U^i(.,f^*)$ satisfies the following optimality equation:\\
	$U^i(p,f^*) = R_i(p,f^*) + \beta \sum_{x \in P} U^i(x,f^*)Q(x\mid p,f^*)$ for all $p \in P$ and $i \in N$.

	\section{\textbf{Related Work}} \label{sec:related}
	
	Inoue et al.~\cite{inoue1997secure} present an implementation example of a secure mobile system which employs a secure mobile IP protocol on stationary security gateways and mobile hosts by modifying IETF standard Mobile IP protocol with IP security primitives, which control the packet flow from a mobile host through multiple security gateways.
	
	Braun et al.~\cite{braun2001secure} describe a solution called secure mobile IP (SecMIP) to provide mobile IP users secure access to their company's firewall protected virtual private network by making a slight adaptation of the end system communication software in order to adapt the mobile IP and IP security protocol implementations to each other.
	
	Leung~\cite{leung2004mobile} provides methods and apparatus for authenticating a mobile node by configuring the server to provide a plurality of security associations associated with a plurality of mobile nodes.
	
	Zao et al.~\cite{zao1999public} present the design and the implementation of a public key management system called Mobile IP Security (MoIPS) built upon a DNS based X.509 Public Key Infrastructure and the innovation in cross certification and zero‐message key generation that can be used with IETF basic and route optimized Mobile IP.
	
	Yokote et al.~\cite{yokote2002intelligent} present a solution to asynchronous security association between nodes by implementing a security association policy server for IPsec in third generation and beyond wireless mobile access, Internet protocol-based digital networks supporting Mobile IP.
	
	Deng et al.~\cite{deng2002defending} point out the weaknesses of two solutions proposed by the IETF Mobile IP Working Group and present a new protocol for securing binding update messages in order to defend against redirection attack.
	
	Hossain et al.~\cite{hossain2011security} explain with illustrative examples major security threats and several existing security solutions on various components of the network involving the mobility and identified additional security holes of these existing solutions and propose some simple mechanisms to counter them.
	
	Lin et al.~\cite{linstochastic} propose Stochastic Game Nets (SGN) to model and deal with the game issues, which takes advantages of both stochastic game theory and Stochastic Petri Nets by inheriting the flexible modeling approach of Stochastic Petri Nets. They also apply the SGN method to model and analyze the network attacks, compute the Nash Equilibrium and best-response strategies to defend the attacks.
	A number of researchers~\cite{michiardi2002game, alpcan2003game, agah2004game, lye2005game, agah2007preventing, luo2010game, roy2010survey, liang2013game, manshaei2013game} have proposed game-theory based solutions for network security problems. However, Lin et al.~\cite{linstochastic} propose Stochastic Game Nets (SGN) to model and deal with the game issues, which takes advantages of both stochastic game theory and Stochastic Petri Nets by inheriting the flexible modeling approach of Stochastic Petri Nets. They also apply the SGN method to model and analyze the network attacks, compute the Nash Equilibrium and best-response strategies to defend the attacks. Wang et al.~\cite{wang2012stochastic, wang2013modeling} later extended this work by applying it to the security analysis for enterprise networks. Our work extends the work of ~\cite{wang2012stochastic, wang2013modeling} but differs in the fact that while they applied Stochastic Game Net (SGN) to model attacks in enterprise networks, we apply Stochastic Game Petri Net (SGPN) to model attacks and defenses in Mobile IP.
	
	\section{\textbf{SGPN Representation}}
\begin{figure}[!htbp]
		\begin{center}
			\includegraphics[width=\columnwidth]{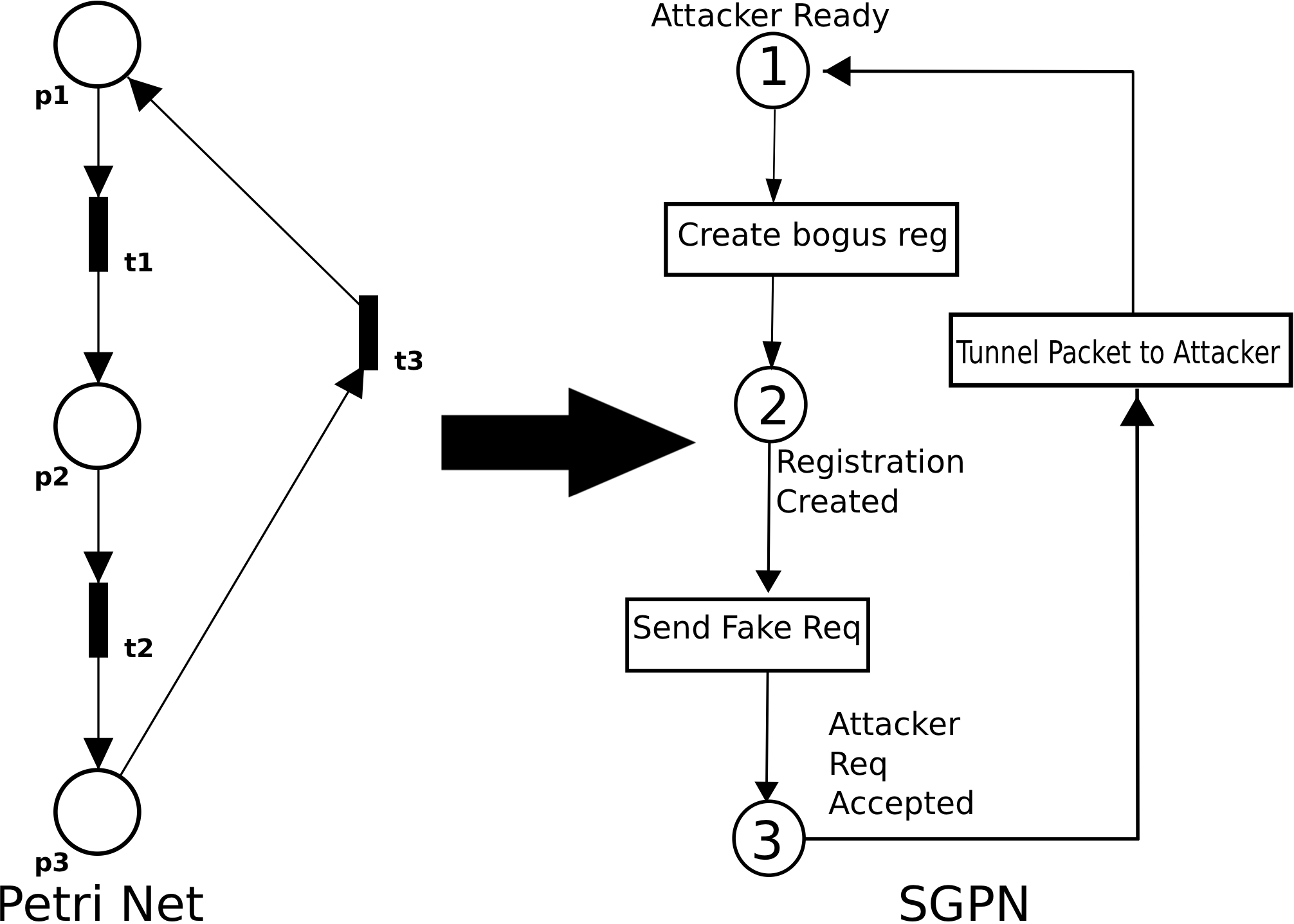}
			\caption{Conversion of SGPN from Petri Net}
			\label{fig:construction}
		\end{center}
	\end{figure}

	We represent our SGPN models according to the following graphical rules: Each place in the Petri Net is represented by the round red colored circles with having the label inside the circle. Each transition in the Petri Net is represented by rectangular red colored shapes with having label placed inside the rectangle. All the arcs are represented by the directed arrows resembling the actual arcs. Figure~\ref{fig:construction} shows an example how the Petri Nets are converted in the SGPN.
	
	\section{\textbf{Attack Modeling}}
	\label{sec:attack}
	
	\subsection{\bf{Denial of Service (DoS) Attack}}
	Denial-of-Service (DoS) is an attack that makes a device or web resource temporarily or indefinitely inaccessible to its authorized users. In Mobile IPv4, DoS attacks is performed by the preclusion of packets from flowing between two nodes. DoS attacks are very common in target sites or services hosted on high-profile web servers such as business sites, credit card payment gateways, banks, and even root name servers. DoS attack is also possible in IPv6. Since the IPv6 support is on par with the IPv4-based feature set, attacks can be carried out over IPv4, and by shifting over to IPv6 it is possible to bypass the defenses that only inspect IPv4 traffic. Generally saying, DoS attack takes the following steps:
	
	\begin{itemize}
		\item In order to initiate the attack, the attacker stays on the path between two nodes to perform the preclusion of packets flowing between them by intercepting the communication between the two nodes directly.
		\item When a mobile node is connected on the foreign network, it must use the registration request to inform its home agent of its current care-of address. Home agent intercepts and tunnels all the traffic destined to mobile node’s home address to its Care-of-Address (CoA).
		\item During the attack, the attacker creates a bogus Registration Request, specifying his own IP address as the CoA for the mobile node.
		\item If the mobile node's home address is fooled by this fake registration request, all packets would be tunneled to the attacker instead of mobile node's actual CoA. Thus, the connection to the mobile node is lost.
	\end{itemize}
	
	\begin{figure}
		\begin{center}
			\includegraphics[width=2.3in, height=3.3in]{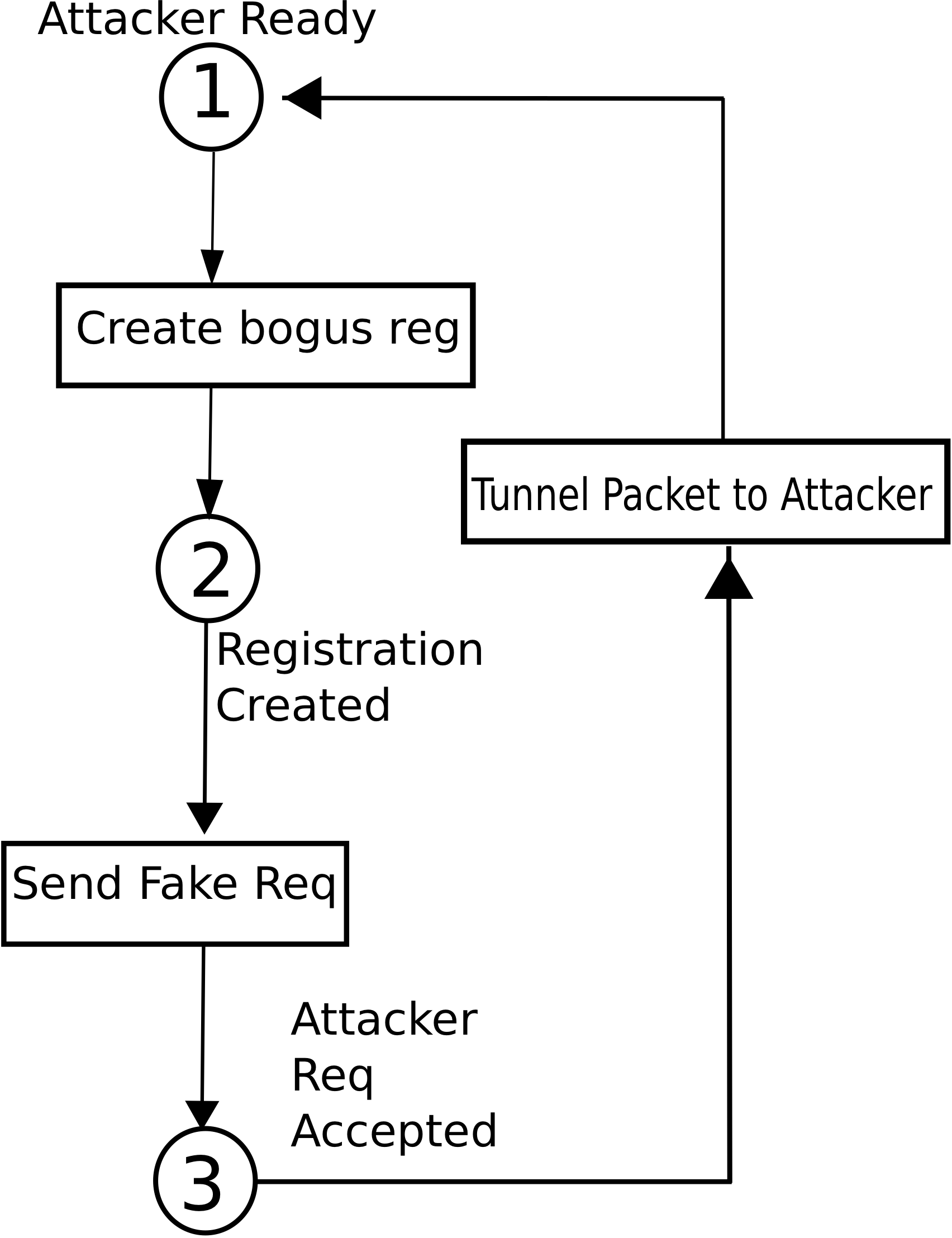}
			\caption{DoS Attack}
			\label{fig:DoS Attack}
		\end{center}
	\end{figure}
	
	\begin{table}
		\centering
		\resizebox{\columnwidth}{!}{%
			\begin{tabular}{|l|l|}
				\hline
				\rowcolor{Gray}\textbf{Place} & \textbf{Description} \\
				\hline
				\textbf{State 1} & Attacker is ready to attack. \\
				\hline
				\textbf{State 2} & Attacker has created bogus registration. \\
				\hline
				\textbf{State 3} & Attacker's fake request is accepted by MN. \\
				\hline
				\rowcolor{Gray}\textbf{Transition} & \textbf{Description} \\
				\hline
				\textbf{Create bogus reg} & Attacker is creating a bogus registration. \\
				\hline
				\textbf{Send Fake Req} & Attacker is sending the fake registration request to MN. \\
				\hline
				\textbf{Tunnel Packet to Attacker} & MN is fooled by the fake request.\\& Data is tunneled to attacker instead of MN. \\
				\hline
			\end{tabular}
		}
		\vspace{5pt}
		\caption{DoS attack place and transition description}
		\label{table:dos:attack}
	\end{table}

	

	\subsection{ \bf{Redirection Attack}}
	Redirection attack is a type of attack in which the intended traffic for the MN is redirected by the attacker through sending a fabricated BU, thus depriving MN from getting data. To launch the redirection attacks, the IP addresses of the communicating nodes has to be known by the attacker. Hence, nodes with well-known IP addresses, such as file servers, public servers and DNS servers are more vulnerable to such attacks. The attack takes place in the following steps:
	
	\begin{itemize}
		
		\item The attacker sends a fake binding update message to CN claiming that the MN has changed its care-of address due to its movement to a new location. 
		\item If the BU is not authenticated, it will be accepted by the CN. CN will now start sending packets to the new CoA which is fake and the MN will not get any traffic. 
		
	\end{itemize}
	
	\begin{figure}
		\begin{center}
			\includegraphics[width=3in]{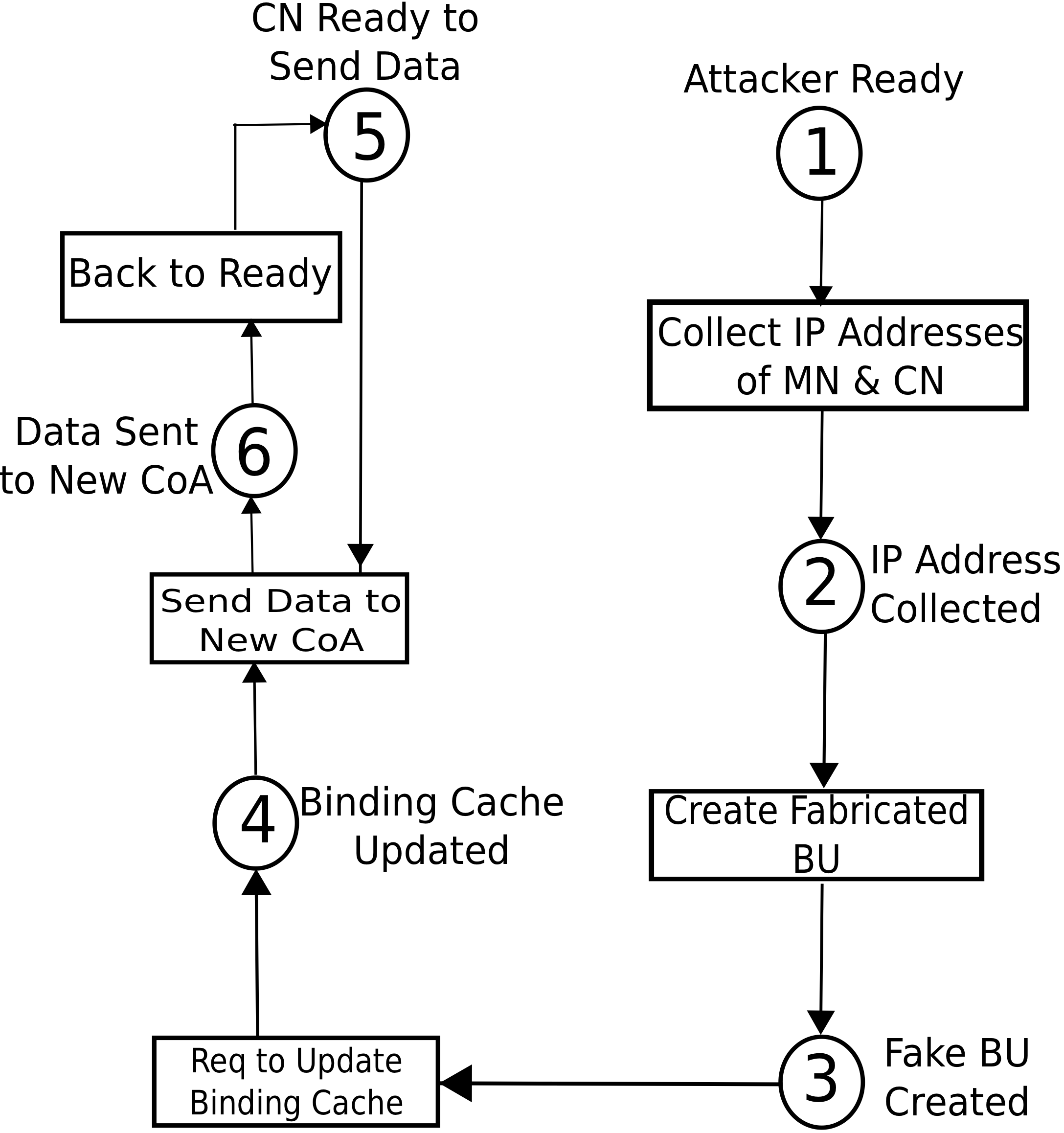}
			\caption{Redirection Attack}
			\label{fig:Redirection Attack}
		\end{center}
	\end{figure}
	
	\begin{table}
		\centering
		\resizebox{\columnwidth}{!}{%
			\begin{tabular}{|l|l|}
				\hline
				\rowcolor{Gray}\textbf{Place} & \textbf{Description} \\
				\hline
				\textbf{State 1} & Attacker is ready. \\
				\hline
				\textbf{State 2} & Attacker has collected IP addresses of MN and CN. \\
				\hline
				\textbf{State 3} & Attacker has created fabricated BU. \\
				\hline
				\textbf{State 4} & CN has updated binding cache using wrong IP address. \\
				\hline
				\textbf{State 5} & CN is ready to send data. \\
				\hline
				\textbf{State 6} & CN has sent data to wrong CoA. \\
				\hline
				\rowcolor{Gray}\textbf{Transition} & \textbf{Description} \\
				\hline
				\textbf{Collect IP Addresses of MN \& CN} & Attacker is collecting IP addresses of MN \& CN. \\
				\hline
				\textbf{Create Fabricated BU} & Attacker is creating fabricated BU. \\
				\hline
				\textbf{Req to Update Binding Cache} & Attacker is requesting to update the binding cache with\\& it's fake BU. \\
				\hline
				\textbf{Send Data to New CoA} & CN is sending data to wrong CoA. \\
				\hline
				\textbf{Back to Ready} & CN is getting back to ready to send data again. \\
				\hline
			\end{tabular}
		}
		\vspace{5pt}
		\caption{Redirection attack place and transition description}
		\label{table:redirection:attack}
	\end{table}
	
	
	

	\subsection{ \bf{Bombing Attack}}
	Bombing attack is an attack where large amount of unwanted data traffic is redirected to the victim node to degrade its performance as well as bandwidth wastage. The attacker may exploit real-time streaming servers for this kind of attack. The attack is performed through the following steps:
	
	\begin{itemize}
		\item First, the attacker establishes a connection with streaming server, and request to download a large stream of data. 
		\item Once the server accepts the request, the attacker starts to get the data along with the sequence number of the data packets. After getting the initial sequence number, the attacker might claim that it has moved to a new location. The attacker uses the IP address of the victim MN in the binding update. 
		\item As a result, subsequent packets from the server will be directed to the victim node causing its performance degradation and bandwidth wastage of the MN. 
	\end{itemize}
	
	\begin{figure}
		\begin{center}
			\includegraphics[width=2.8in]{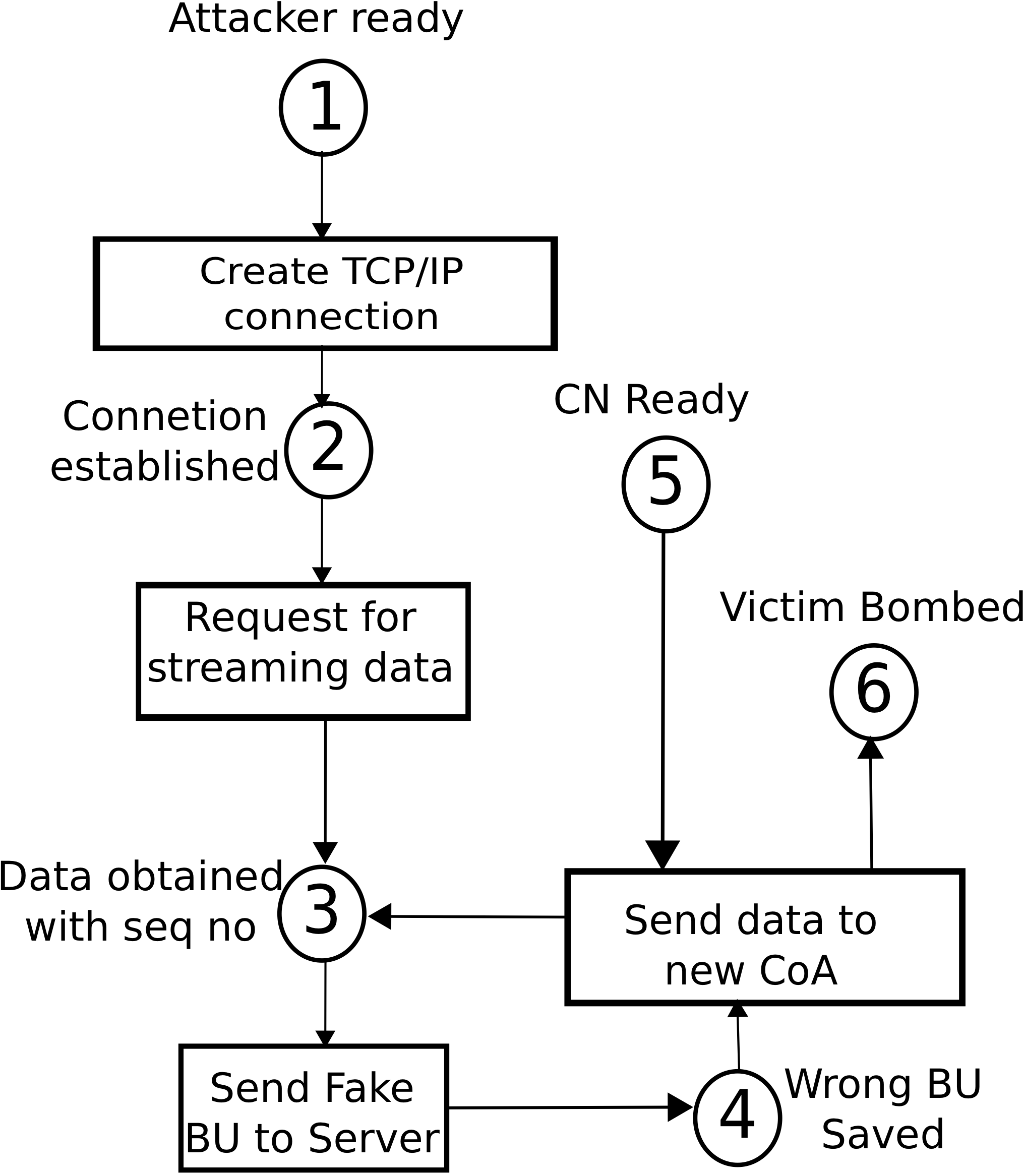}
			\caption{Bombing Attack}
			\label{fig:Bombing Attack}
		\end{center}
	\end{figure}
	
	\begin{table}
		\centering
		\resizebox{\columnwidth}{!}{%
			\begin{tabular}{|l|l|}
				\hline
				\rowcolor{Gray}\textbf{Place} & \textbf{Description} \\
				\hline
				\textbf{State 1} & Attacker is ready to attack. \\
				\hline
				\textbf{State 2} & Attacker has established TCP connection with streaming server. \\
				\hline
				\textbf{State 3} & Attacker has obtained data packets from streaming server along with\\& sequence numbers. \\
				\hline
				\textbf{State 4} & CN updates the binding cache with wrong BU. \\
				\hline
				\textbf{State 5} & CN (Streaming Server) is ready to send data. \\
				\hline
				\textbf{State 6} & Victim MN has received unsolicited stream of data from streaming server. \\
				\hline
				\rowcolor{Gray}\textbf{Transition} & \textbf{Description} \\
				\hline
				\textbf{Create TCP/IP Connection} & Attacker is creating a TCP/IP connection with server. \\
				\hline
				\textbf{Request for streaming data} & Attacker is requesting for streaming data from streaming server. \\
				\hline
				\textbf{Send fake BU to Server} & Attacker is sending fake BU to server specifying that it has changed\\& its location. \\
				\hline
				\textbf{Send data to new CoA} & CN, in this case the streaming server is sending data to victim's IP. \\
				\hline
			\end{tabular}
		}
		\vspace{5pt}
		\caption{Bombing attack place and transition description}
		\label{table:bombing:attack}
	\end{table}
	
	
	
	
	\subsection{ \bf{Replay Attack}}
	Replay attack is a type of attack that takes the advantage of a previously recorded binding update by replaying it when the victim (MN) moves to some new location, thereby disrupting the communication between the CN and the MN. Replay attack works by the following steps:
	
	\begin{itemize}
		\item The attacker stays in close proximity to the MN or CN, for example being in the same radio access network. Thus, the attacker has the ability to record any BU send by the MN to the CN. 
		\item When MN moves to a new network, it sends a BU to the CN.  The attacker listening to such BU records the BU to use for replay attack in future. 
		\item When the MN moves to another new location, the attacker replays the recorded BU to the CN to trick CN. If the CN accepts such replay message, it would start sending packets to the old address thinking that MN has again moved to the old address. Thus, traffic from CN are redirected to a non-existing IP-address, thereby interrupting the communication to the MN.
	\end{itemize}
	
	\begin{figure}
		\begin{center}
			\includegraphics[width=\columnwidth]{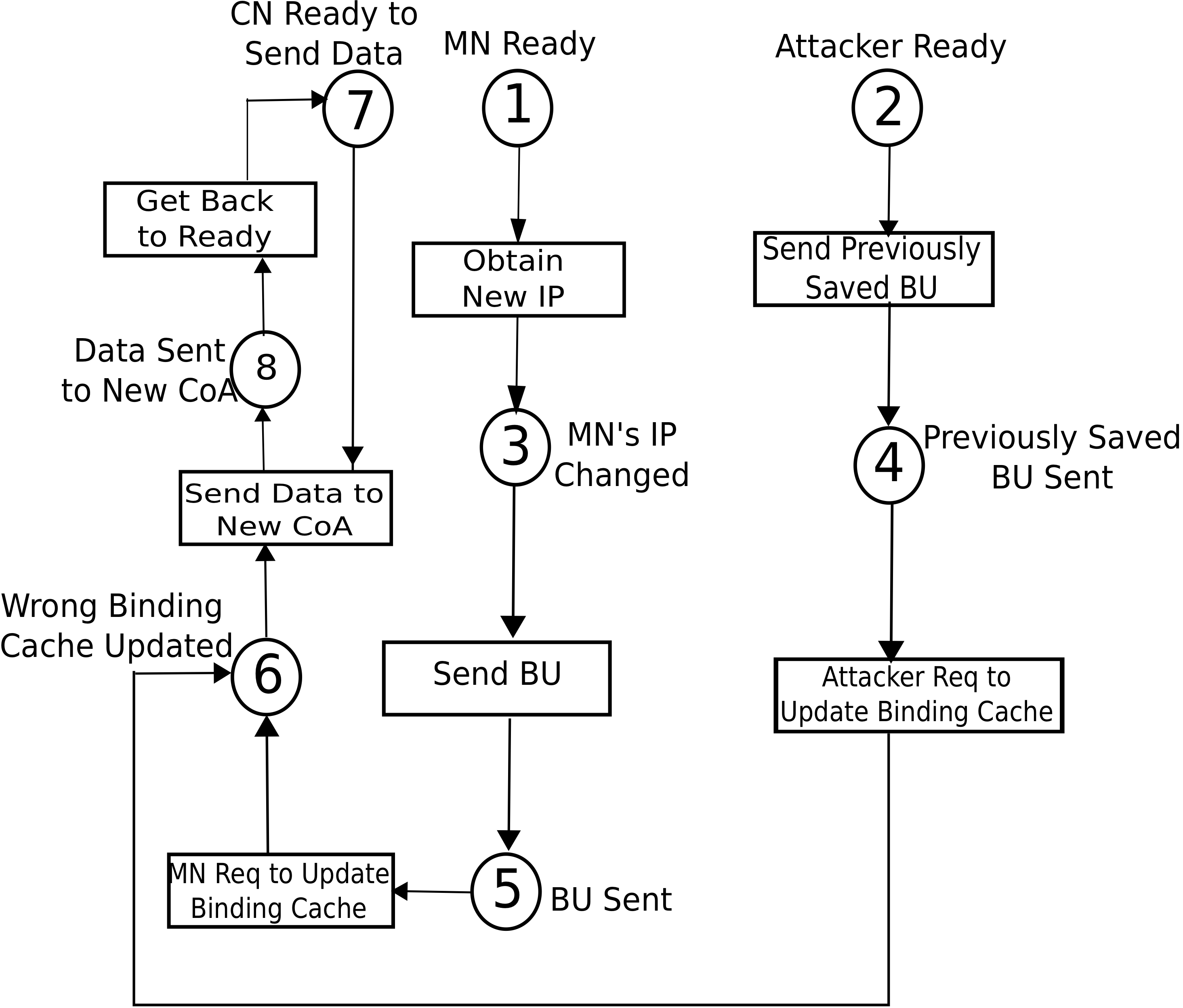}
			\caption{Replay Attack}
			\label{fig:Replay Attack}
		\end{center}
	\end{figure}
	
	\begin{table}
		\centering
		\resizebox{\columnwidth}{!}{%
			\begin{tabular}{|l|l|}
				\hline
				\rowcolor{Gray}\textbf{Place} & \textbf{Description} \\
				\hline
				\textbf{State 1} & MN is ready. \\
				\hline
				\textbf{State 2} & Attacker is ready. \\
				\hline
				\textbf{State 3} & MN's IP is changed due to change of its location. \\
				\hline
				\textbf{State 4} & Attacker has sent a fake BU which was recorded before. \\
				\hline
				\textbf{State 5} & MN has sent the BU to the CN. \\
				\hline
				\textbf{State 6} & CN has updated the binding cache with attacker's BU. \\
				\hline
				\textbf{State 7} & CN is ready to send data. \\
				\hline
				\textbf{State 8} & CN has sent data to updated wrong CoA. \\
				\hline
				\rowcolor{Gray}\textbf{Transition} & \textbf{Description} \\
				\hline
				\textbf{MN ready} & MN is becoming ready to interact with CN. \\
				\hline
				\textbf{Obtain new IP} & MN is obtaining a new IP because it has changed its location. \\
				\hline
				\textbf{Send Previously Saved BU} & MN is sending BU to CN. \\
				\hline
				\textbf{MN req to update BU} & MN is requesting to update binding cache with its BU. \\
				\hline
				\textbf{Attacker req to update BU} & Attacker is requesting to update binding cache with its BU. \\
				\hline
				\textbf{Send data to new CoA} & CN is sending data to fake CoA. \\
				\hline
				\textbf{Get Back to Ready} & CN is getting ready to send data again. \\
				\hline
			\end{tabular}
		}
		\vspace{5pt}
		\caption{Replay attack place and transition description}
		\label{table:replay:attack}
	\end{table}
	
	
	

	\section{\textbf{Defense Modeling}}
	\label{sec:defense}
	
	\subsection{ \bf{Denial of Service (DoS) Defense}}
	The defense for the DoS attack is simply to prevent bogus registrations by the attacker. This can be done by imposing strong authentication on in all registration traffic exchange between a mobile node and its home IP agent. Under the assumption that the shared secret key is protected, this can insure that the traffic intended for mobile node is inaccessible to the attacker. All implementation of Mobile IP supports the default algorithm MD5 for authentication, however mobile node and home agent can use any authentication algorithm they agree upon.
	
	\begin{figure}
		\begin{center}
			\includegraphics[width=\columnwidth]{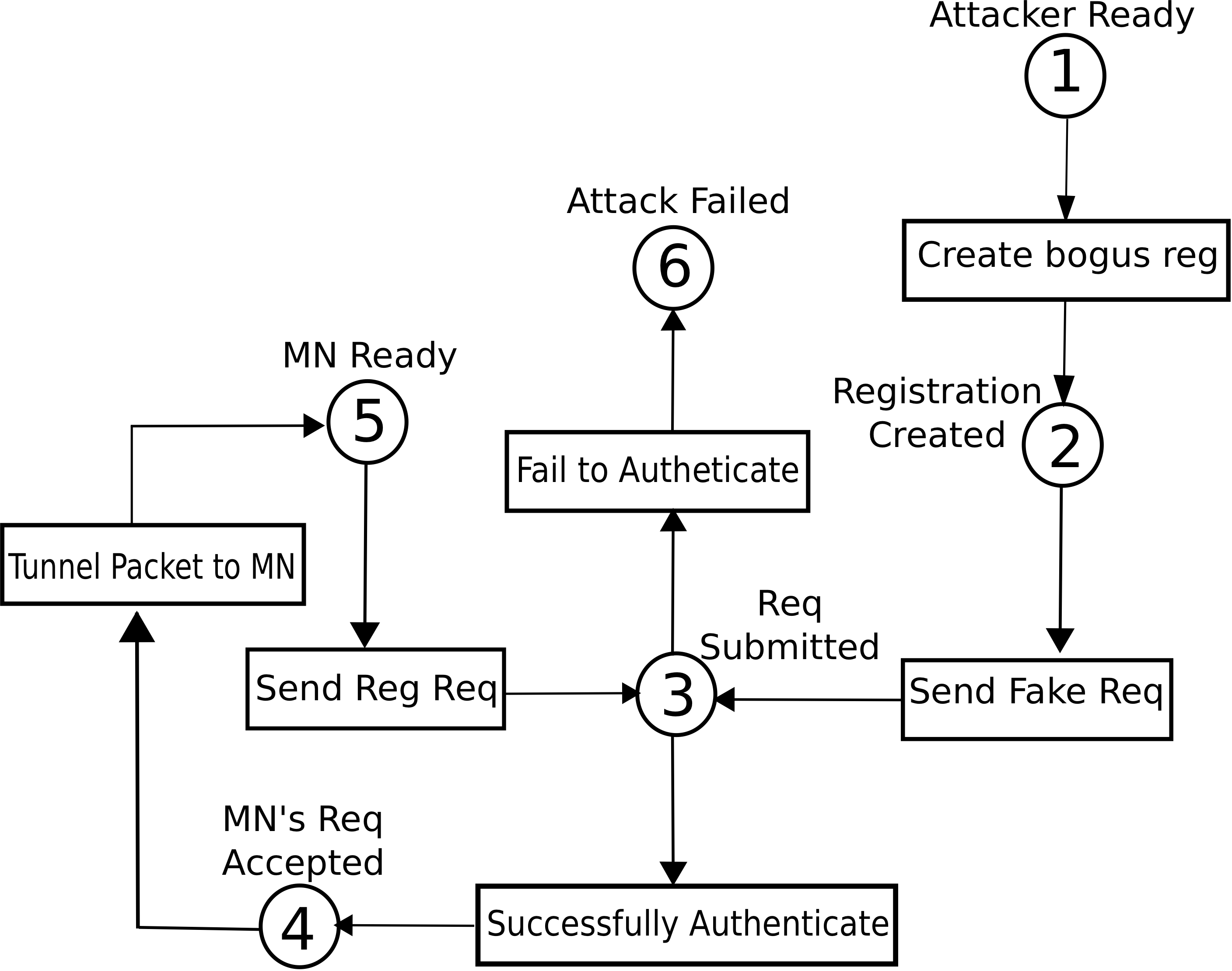}
			\caption{DoS Defense}
			\label{fig:DoS Solution}
		\end{center}
	\end{figure}
	
	\begin{table}
		\centering
		\resizebox{\columnwidth}{!}{%
			\begin{tabular}{|l|l|}
				\hline
				\rowcolor{Gray}\textbf{Place} & \textbf{Description} \\
				\hline
				\textbf{State 1} & Attacker is ready to attack. \\
				\hline
				\textbf{State 2} & Attacker has created bogus registration. \\
				\hline
				\textbf{State 3} & Attacker's fake registration request is submitted to CN. \\
				\hline
				\textbf{State 4} & MN's authenticate request is accepted by CN. \\
				\hline
				\textbf{State 5} & MN is ready to send registration request. \\
				\hline
				\textbf{State 6} & Attacker's authentication is failed and attack is not done. \\
				\hline
				\rowcolor{Gray}\textbf{Transition} & \textbf{Description} \\
				\hline
				\textbf{Create bogus reg} & Attacker is creating a bogus registration. \\
				\hline
				\textbf{Send Fake Req} & Attacker is sending the fake registration request to MN. \\
				\hline
				\textbf{Send Reg Req} & MN is sending valid registration request to CN. \\
				\hline
				\textbf{Successfully Authenticate} & CN is successfully authenticating the registration request of the MN. \\
				\hline
				\textbf{Tunnel Packet to MN} & CN is tunneling packet to MN. \\
				\hline
				\textbf{Fail to Authenticate} & CN is unsuccessfully authenticating the registration request\\& of the attacker. \\
				\hline
			\end{tabular}
		}
		\vspace{5pt}
		\caption{DoS defense place and transition description}
		\label{table:dos:defense}
	\end{table}

	
	
	
	\subsection{ \bf{Redirection Defense}}
	In order to defend against the redirection attack, CN should authenticate the BU before updating it. CN should send data to the new location only when the BU is authenticated. Another mitigation to this problem could be frequently changing the IP address by the communicating nodes. However, this is impractical since this additional security mechanism will make the mobility protocol slower and more complex.
	Brief description about the states and the transitions for solution scenario of redirection attack are given below-
	
	\begin{figure}
		\begin{center}
			\includegraphics[width=2.8in]{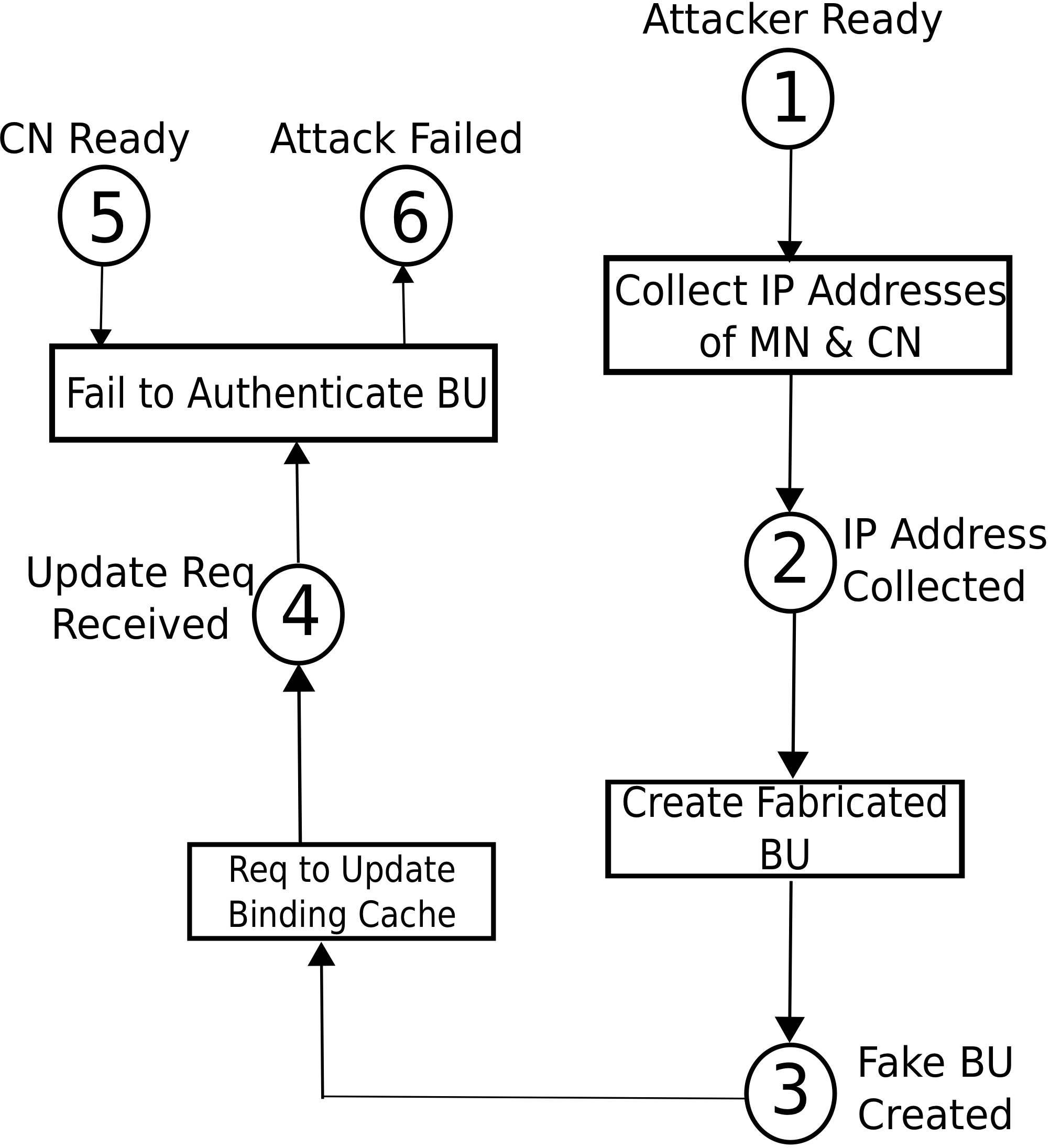}
			\caption{Redirection Defense}
			\label{fig:Redirection Solution}
		\end{center}
	\end{figure}
	
	\begin{table}
		\centering
		\resizebox{\columnwidth}{!}{%
			\begin{tabular}{|l|l|}
				\hline
				\rowcolor{Gray}\textbf{Place} & \textbf{Description} \\
				\hline
				\textbf{State 1} & Attacker is ready. \\
				\hline
				\textbf{State 2} & Attacker has collected IP addresses of MN and CN. \\
				\hline
				\textbf{State 3} & Attacker has created fabricated BU. \\
				\hline
				\textbf{State 4} & CN has received requests to update binding cache. \\
				\hline
				\textbf{State 5} & CN is ready to send data. \\
				\hline
				\textbf{State 6} & CN has failed to authenticate wrong BU and attack is failed. \\
				\hline
				\rowcolor{Gray}\textbf{Transition} & \textbf{Description} \\
				\hline
				\textbf{Collect IP Addresses of MN \& CN} & Attacker is collecting IP addresses of MN \& CN. \\
				\hline
				\textbf{Create Fabricated BU} & Attacker is creating fabricated BU. \\
				\hline
				\textbf{Req to Update Binding Cache} & Attacker is requesting to update the binding cache with\\& it's fake BU. \\
				\hline
				\textbf{Fail to Authenticate BU} & CN is failing to authenticate attacker's fake BU. \\
				\hline
			\end{tabular}
		}
		\vspace{5pt}
		\caption{Redirection defense place and transition description}
		\label{table:redirection:defense}
	\end{table}
	
	
	
	
	\subsection{ \bf{Bombing Defense}}
	For defending against bombing attack, server can send a hello packet to the new location and wait for the acknowledgement. After receiving acknowledgement from the new location, server will send the desired data to MN. Thus, bombing attack can be prevented. However, one possible problem of this defense is that, the attacker can spoof acknowledgement to the server as it knows the initial sequence number making a continuous flow of data streams sent to the victim. One possible solution of this could be to use the TCP RESET signal by the victim node to immediately stop such flow of data stream.
	
	\begin{figure}
		\begin{center}
			\includegraphics[width=3in]{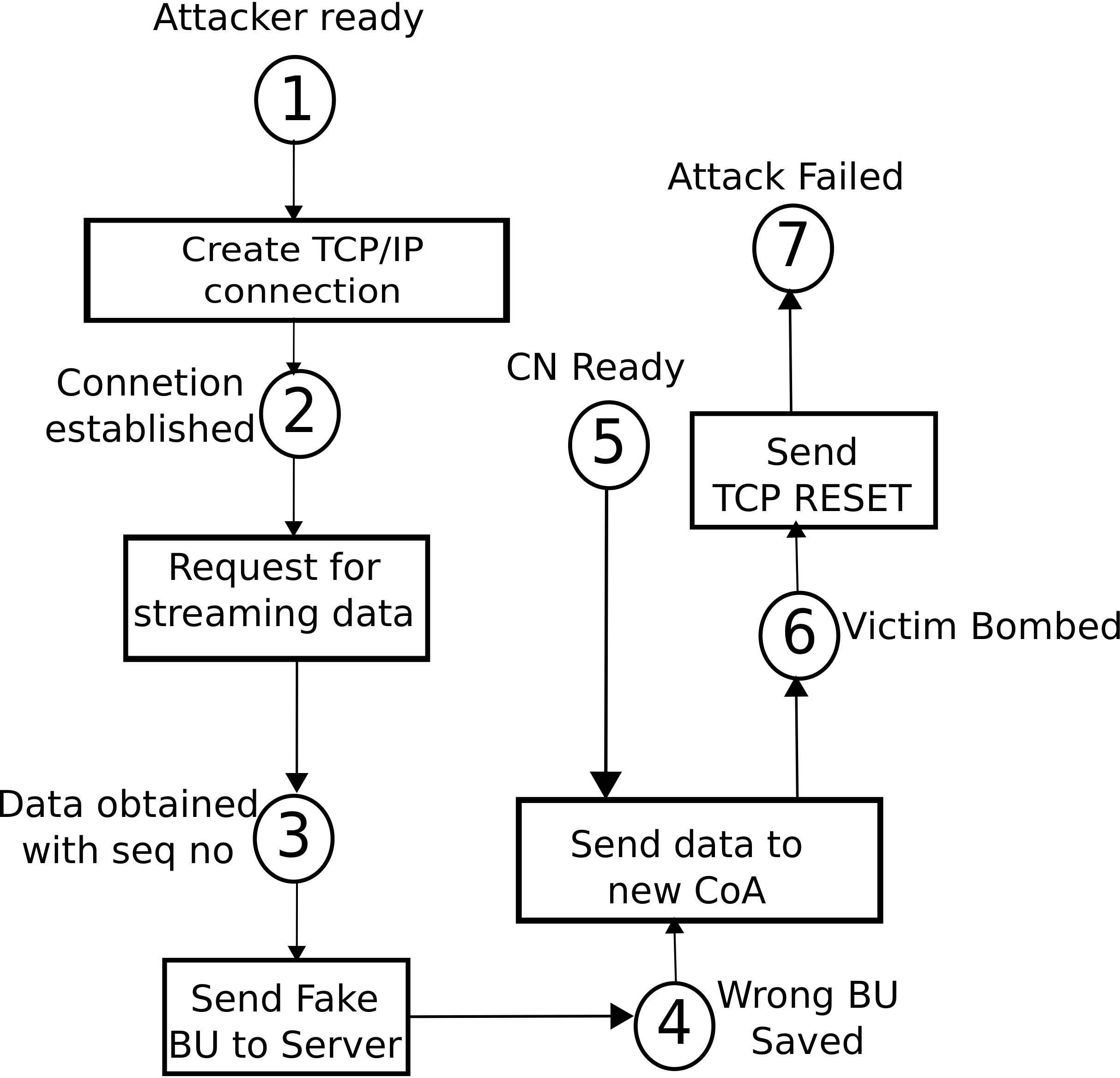}
			\caption{Bombing Defense}
			\label{fig:Bombing Solution}
		\end{center}
	\end{figure}
	
	\begin{table}
		\centering
		\resizebox{\columnwidth}{!}{%
			\begin{tabular}{|l|l|}
				\hline
				\rowcolor{Gray}\textbf{Place} & \textbf{Description} \\
				\hline
				\textbf{State 1} & Attacker is ready to attack. \\
				\hline
				\textbf{State 2} & Attacker has established TCP connection with streaming server. \\
				\hline
				\textbf{State 3} & Attacker has obtained data packets from streaming server along with\\& sequence numbers. \\
				\hline
				\textbf{State 4} & CN updates the binding cache with wrong BU. \\
				\hline
				\textbf{State 5} & CN (Streaming Server) is ready to send data. \\
				\hline
				\textbf{State 6} & Victim MN has received unsolicited stream of data from streaming server. \\
				\hline
				\textbf{State 7} & Victim MN has sent TCP RESET and attack is failed. \\
				\hline
				\rowcolor{Gray}\textbf{Transition} & \textbf{Description} \\
				\hline
				\textbf{Create TCP/IP Connection} & Attacker is creating a TCP/IP connection with server. \\
				\hline
				\textbf{Request for streaming data} & Attacker is requesting for streaming data from streaming server. \\
				\hline
				\textbf{Send fake BU to Server} & Attacker is sending fake BU to server specifying that it has changed\\& its location. \\
				\hline
				\textbf{Send data to new CoA} & CN, in this case the streaming server is sending data to victim's IP. \\
				\hline
				\textbf{Send TCP RESET} & Victim MN is sending TCP RESET signal to CN. \\
				\hline
			\end{tabular}
		}
		\vspace{5pt}
		\caption{Bombing defense place and transition description}
		\label{table:bombing:defense}
	\end{table}

	
	
	
	\subsection{ \bf{Replay Defense}}
	In order to defend against the replay attack, CN should authenticate the BU before updating it. However, it is very difficult to defend against replay attack since the attacker is already using some authenticated BU. On possible solution for this problem is to use the sequence number as the authentication parameter. CN should store the sequence number of the previously send binding updates in a stable storage. CN should send data to the new location only when the BU is authenticated and the sequence number is not repeated.
	
	\begin{figure}
		\begin{center}
			\includegraphics[width=\columnwidth]{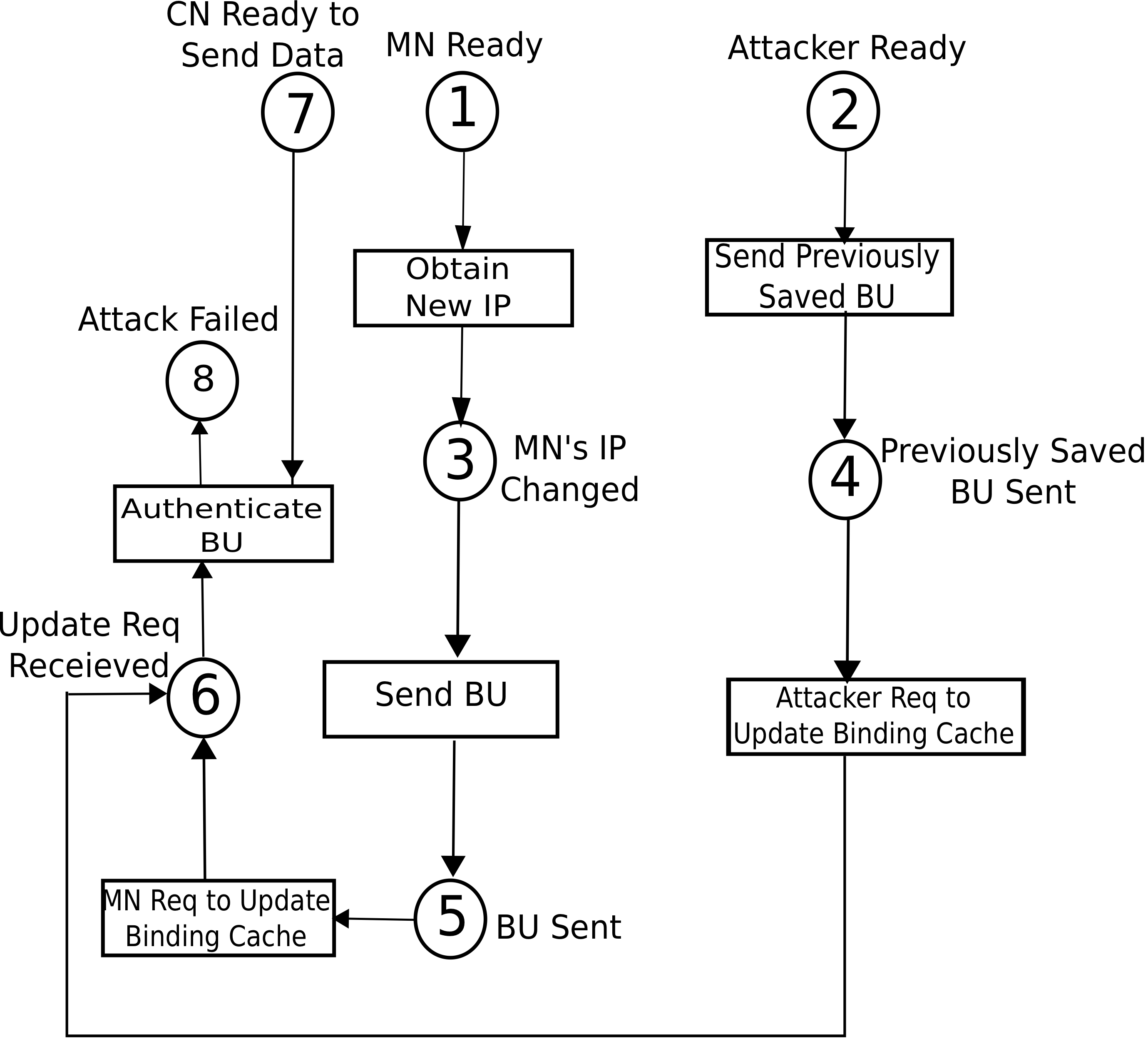}
			\caption{Replay Defense}
			\label{fig:Replay Defense}
		\end{center}
	\end{figure}
	
	\begin{table}
		\centering
		\resizebox{\columnwidth}{!}{%
			\begin{tabular}{|l|l|}
				\hline
				\rowcolor{Gray}\textbf{Place} & \textbf{Description} \\
				\hline
				\textbf{State 1} & MN is ready. \\
				\hline
				\textbf{State 2} & Attacker is ready. \\
				\hline
				\textbf{State 3} & MN's IP is changed due to change of its location. \\
				\hline
				\textbf{State 4} & Attacker has sent a fake BU which was recorded before. \\
				\hline
				\textbf{State 5} & MN has sent the BU to the CN. \\
				\hline
				\textbf{State 6} & CN has received requests to update the binding cache. \\
				\hline
				\textbf{State 7} & CN is ready to send data. \\
				\hline
				\textbf{State 8} & CN has failed to authenticate fake BU and attack is failed. \\
				\hline
				\rowcolor{Gray}\textbf{Transition} & \textbf{Description} \\
				\hline
				\textbf{MN ready} & MN is becoming ready to interact with CN. \\
				\hline
				\textbf{Obtain new IP} & MN is obtaining a new IP because it has changed its location. \\
				\hline
				\textbf{Send Previously Saved BU} & MN is sending BU to CN. \\
				\hline
				\textbf{MN req to update BU} & MN is requesting to update binding cache with its BU. \\
				\hline
				\textbf{Attacker req to update BU} & Attacker is requesting to update binding cache with its BU. \\
				\hline
				\textbf{Authenticate BU} & CN is authenticating the BU send by the attacker. \\
				\hline
			\end{tabular}
		}
		\vspace{5pt}
		\caption{Replay defense place and transition description}
		\label{table:replay:defense}
		\vspace{-15pt}
	\end{table}
	
	
	

	\section{\textbf{Evaluation}}
	\label{sec:user:evaluation}
	
	We now evaluate our modeling approach by calculating the steady state probabilities of the attack and defense scenarios. We compare the infection probability of the attacks and our proposed defense scenarios. We use simulation technique using MATLAB to get the probability values and for plotting the simulation data. We use algorithm~\ref{alg} to calculate the steady state probability from the Nash Equilibrium. 
	
	\begin{algorithm}
		\caption{Evaluate SGPN Model}\label{alg}
		\begin{algorithmic}[1]
			
			\State Let, $\bm{\{A_a^1,A_a^2\}, \{A_n^1,A_n^2\}, \{D_d^1,D_d^2\}}$ and $\bm{\{D_n^1,D_n^2\}}$ be the reward values for the attacker's attack, attacker's not attack, defender's defend and defender's not defend actions respectively.
			\State $P_A \gets \textit{Attacker's probability of Attacking}$
			\State $P_D \gets \textit{Defender's probability of defending}$
			\State Calculate $\bm{NE \gets \{P_A,P_D\}}$ by solving the following:
			\State $P_A\times A_a^1 + (1-P_A)\times A_n^1 = P_A\times D_d^1 + (1-P_A)\times D_n^1$
			\State $P_D\times A_a^2 + (1-P_D)\times D_d^2 = P_D\times A_n^2 + (1-P_D)\times D_n^2$
			\State $M_r \gets \textit{Reduced attack-defend model}$
			\State Generate \textit{reachability tree,} $\bm T_r$ from the \textit{attack-defend model,} $\bm M_r$
			\State Calculate \textit{steady state probability,} $\bm \pi$ using $\bm T_r$ and ${\bm NE}$
			\State We can say that, if defender defends $\bm P_D$\% of the time, the probability of a successful attack is $\bm \pi$\%
		\end{algorithmic}
	\end{algorithm}
	
\begin{figure}
		\begin{center}
			\includegraphics[width=2.5in]{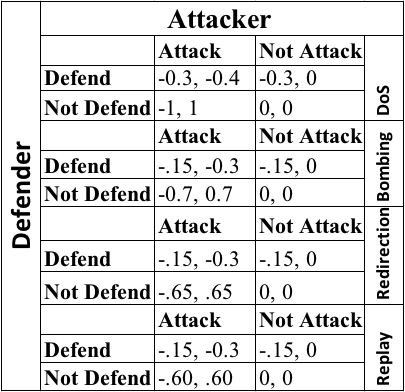}
			\caption{Reward Summary used in attack simulations}
			\label{table:reward}
		\end{center}
	\end{figure}

	In the simulations, we have used the following reward values for the attacker and defender that can be summarized in Table~\ref{table:reward}.
	
	
	\subsection{\textbf{Steady State Probabilities}}
\begin{figure}
		\begin{center}
			\includegraphics[width=3in]{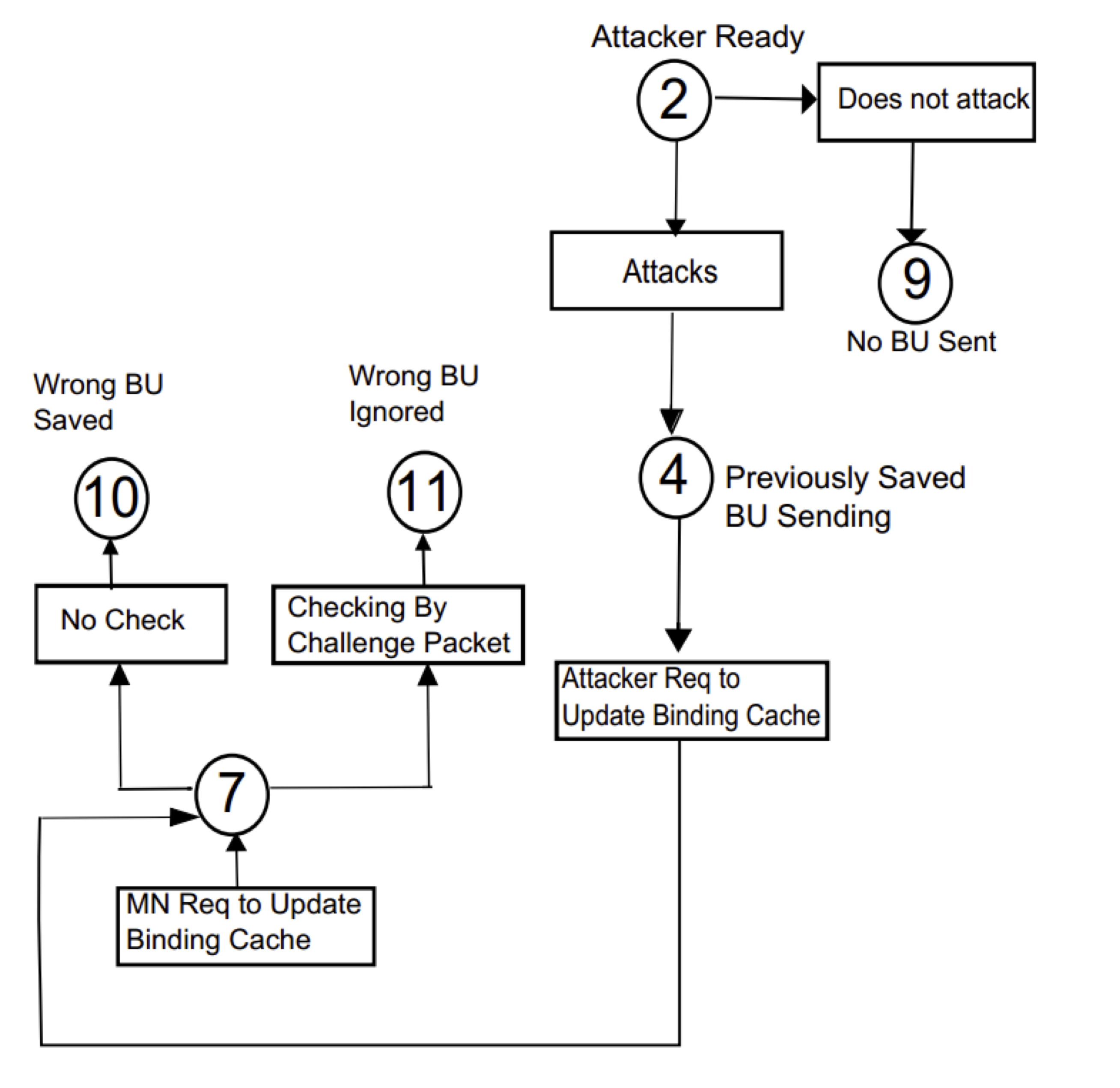}
			\caption{Attack-defend model for Replay Defense}
			\label{fig:attack_defense}
		\end{center}
	\end{figure}

\begin{figure}
		\begin{center}
			\includegraphics[width=2.3in]{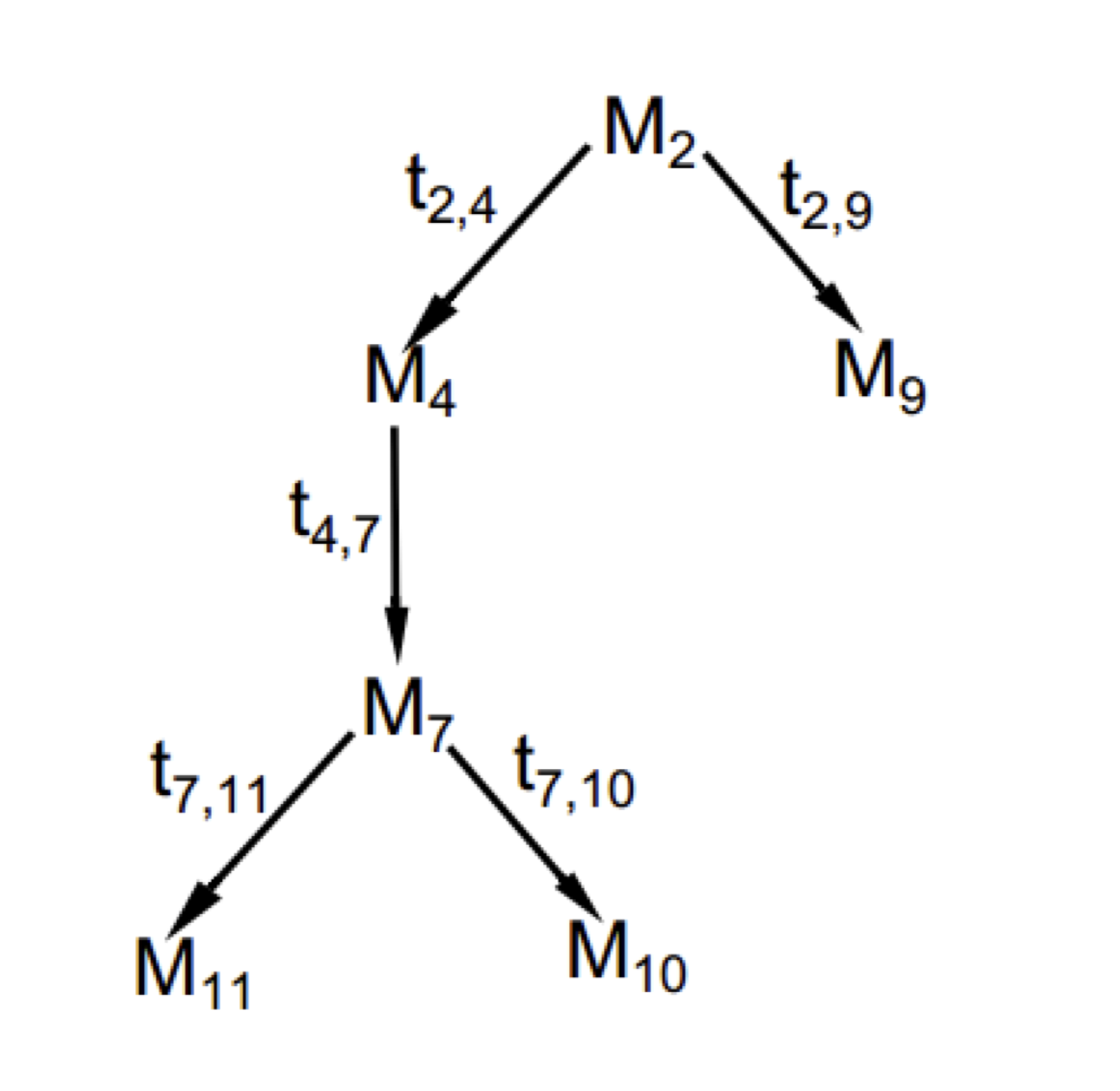}
			\caption{Reachability Tree from attack-defend model}
			\label{fig:reachability}
		\end{center}
	\end{figure}

	We now calculate the steady state probabilities of the attack models. For that we have generated the reachability trees from the models. For the simplicity of calculation, we have reduced the models by keeping the attack-defend states and removing the non attack-defend states. Figure~\ref{fig:attack_defense} shows an example of the reduced attack-defend model of the replay attack defense. Figure~\ref{fig:reachability} shows the reachability tree generated from the attack-defend model of the replay attack defense. 
	
	Using the reward values from Figure~\ref{table:reward} for replay attack, we can calculate the Nash Equilibrium for the replay attack defense. Plugging in the values, $A_a^1 =-0.3 , A_n^1= 0.6, D_d^1= 0, D_n^1 = 0, A_a^2 = -.15, D_d^2= -.15, A_n^2 = -0.6, D_n^2 =  0$, we get the NE = \{0.25, 0.6667\}.
	By using the reachability tree and NE probabilities, we get the following steady state probabilities $\{\pi_9, \pi_{10}, \pi_{11}\} = \{0.75, 0.08333, 0.16667\}$. From the above calculations, we can say that if the defender defends 66.67\% of the time, the probability of a successful attack is only 8.333\%.
	
	In a similar fashion, we calculate the NE and steady state probabilities of the remaining attacks: DoS attack, bombing attack and redirection attack. We find the values of $P_D$ as $0.724$, $0.70$ and $0.684$ respectfully. From these, we find the values of steady state probabilities as $0.0857412$, $0.0642$ and $0.07287$ respectively. Thus, we can conclude that under the optimal strategy, an IDS needs to remain active $72.4\%$, $70\%$, $68.4\%$ and $66.6\%$ of the time to restrict the attacker’s success rate to $8.5\%$, $6.4\%$, $7.2\%$ and $8.3\%$ for the Denial-of-Service (DoS) attack, bombing attack, redirection attack and replay attack respectively.
	
	\subsection{\textbf{Comparison with state-of-the-art approach}}
	Our model performs better than many state-of-the-art approaches for intrusion detection in wireless networks using game theory. Though the other researchers have performed their analysis in a slightly different contexts, we believe our work is comparable to them. For example, Ma et al.~\cite{4570911} has shown that their approach can detect the attacker in 70\% to 85\% of the cases. In other words, the attacker is able to perform successful attacks in 15\% to 30\% cases. In our case, on average the attacker is able to perform successful attack only in 7.6\% of the cases. In other words, IDS can detect the attacker over 92\% of the cases. Similar comparisons can also be drawn with other intrusion detection approaches modeled with game theory. 
	
	\section{\textbf{Conclusions}}
	\label{sec:conclusions}
	
	In our works, SGPN bring together the Game Theory and the Stochastic Petri Nets, and thus takes the gains of both stochastic game theory and Stochastic Petri Nets. It is our strong believe that the proposed SGPN approach can unwrap a new possibility to deal with the security issues in wireless and communication networks. We show that under the optimal strategy, our model can restrict the attacker’s success rate to $8.5\%$, $6.4\%$, $7.2\%$ and $8.3\%$ for the Denial-of-Service (DoS) attack, bombing attack, redirection attack and replay attack respectively. The IDS needs to remain active only $72.4\%$, $70\%$, $68.4\%$ and $66.6\%$ of the time to achieve such performances. Future networks will rely on autonomous and distributed architectures to improve the competence and suppleness of mobile applications, and our SGPN provides the ideal framework for designing efficient and robust distributed algorithms.
	
	\section{\textbf{ACKNOWLEDGMENT}}
	The authors would like to thank Dr. Bogdan Carbunar for his valuable advice and guidance.
	
	\bibliographystyle{IEEEtran}
	\bibliography{ip_mobilty_IEEE}
	
\end{document}